\documentclass[pra,aps,showpacs,twocolumn,superscriptaddress,amsmath,amssymb,floatfix]{revtex4}
\usepackage{graphicx}
 
\newcommand{\be}{\begin{equation}}
\newcommand{\ee}{\end{equation}}

\newcommand{\kk}{{\mathbf k}}

\newcommand{\eq}[1]{(\ref{#1})}
\newcommand{\mume}{\mu \rm m}

\begin{document}
\title{Probing the excitation spectrum of nonresonantly pumped polariton condensates}
\author{Michiel Wouters}
\affiliation{Institute of Theoretical Physics, Ecole Polytechnique
F\'ed\'erale de Lausanne EPFL, CH-1015 Lausanne, Switzerland}
\author{Iacopo Carusotto}
\affiliation{BEC-CNR-INFM and Dipartimento di Fisica, Universit\`a di Trento, 
I-38050 Povo, Italy}
\affiliation{Institute of Quantum Electronics, ETH Z\"urich, 8093 Z\"urich, Switzerland}
\begin{abstract}
We propose a four wave mixing experiment to probe the elementary excitation spectrum of a non-equilibrium Bose-Einstein condensate of exciton-polaritons under non-resonant pumping.  Analytical calculations based on mean-field theory show that this method is able to reveal the characteristic negative energy feature of the Bogoliubov dispersion.  Numerical simulations including the finite spatial profile of the excitation laser spot and a weak disorder confirm the practical utility of the method for realistic condensates.
\end{abstract}

\maketitle

A series of remarkable experiments have recently demonstrated the occurrence of Bose-Einstein condensation in systems of exciton-polaritons in semiconductor microcavities~\cite{kasprzak,yamamoto,snoke,christopoulos,bloch}.
These observations, together with parallel ones on magnon condensation in magnetic solids~\cite{magnons}, are opening the way to the investigation of the Bose-Einstein condensation phase transition and of Bose-condensed quantum fluids in completely new regimes far from thermal equilibrium where the state of the condensate is no longer determined by a thermal equilibrium condition, but by a dynamical balance between driving and dissipation~\cite{non-eq}.

A key quantity in the theory of quantum fluids is the dispersion of the elementary excitations, which determines the dynamics of the system in response to external perturbations and, in particular, plays a central role in determining its superfluidity properties~\cite{superfluidity}.
While a detailed knowledge is nowadays available of the phonon and roton branches of liquid Helium and of the Bogoliubov modes of dilute atomic gases~\cite{BECbook}, not much experimental work has been performed yet on polariton condensates. 
On one hand, pioneering luminescence~\cite{OPO_luminescence} and pump-and-probe experiments~\cite{OPO_PL_pumpprobe,noi_madrid} have addressed the excitations of resonantly pumped condensates in an OPO configuration; on the other hand, luminescence experiments suggesting a linear dispersion of the elementary excitations in nonresonantly excited polariton condensates have been recently reported~\cite{yamamoto_bog}. Still, none of these works has provided complete evidence of the peculiar features that were predicted to appear because of the non-equilibrium nature of polariton condensates~\cite{littlewood,goldstone,nonresonant}.

From a different standpoint, while quite some evidence is available for ``off-branch'' scattering processes in OPO configurations~\cite{OPO_luminescence,OPO_PL_pumpprobe}, no observation of the related negative energy ``ghost'' branch that is expected to appear under non-resonant pumping as a consequence of polariton-polariton interactions has been reported yet. 
Most likely, this stems from the fact that experiments were based on luminescence spectroscopy~\cite{yamamoto_bog}, so that the emission from the ghost branch was easily masked by the much stronger background of the condensate emission.
A few recent theoretical work has proposed more refined schemes that appear suited to overcome this difficulty and detect this elusive branch by looking at either the absorption~\cite{paul} and resonant Rayleigh scattering~\cite{cambridge} spectra, or at the density response to an external perturbation~\cite{sarchi}.

In the present work, we push these ideas forward and we propose a simple four-wave mixing (FWM) spectroscopy scheme to measure the dispersion of all elementary excitations branches of a non-resonantly pumped polariton condensate: 
of the three incident beams needed in a FWM experiment, two are provided by the condensate, and one by an external laser field. The dispersion of excitations is inferred by scanning the energy and wavevector of this latter beam and recording the transmitted, reflected, and/or four-wave mixed beams. 
A first application of related FWM techniques to polariton systems across the parametric oscillation threshold was reported in~\cite{jerome}, but no specific interest was paid to the excitation modes of the condensate itself.

The mean-field Gross-Pitaevskii formalism to describe the polariton condensate dynamics that was introduced in~\cite{nonresonant} is briefly reviewed in Sec.\ref{sec:elex} and then applied in Sec.\ref{sec:FWM} to describe the FWM response of polariton condensates. 
Closed formulas are extracted for the transmission, reflection and FWM signals in the spatially homogeneous case and then used to discuss the main features of the spectra.
Generalization to the experimentally relevant case of finite-size condensate is investigated in Sec.\ref{sec:probe}: the efficiency of the FWM method for realistic condensates is confirmed by means of a numerical solution of the polariton GPE in non-uniform geometries.
Conclusions are finally drawn in Sec. \ref{sec:concl}. 

\section{The elementary excitation spectrum \label{sec:elex}}

As a consequence of the short life time of polaritons in state of the art microcavities, the polariton condensate can hardly be considered as a thermal equilibrium object: Continuous external pumping is necessary to keep the condensate in a stationary state, and the state of the condensate is determined by a dynamical balance between pumping and dissipation.
While the momentum distribution of large systems still appear to follow at large energies the typical exponential law of equilibrium statistical mechanics~\cite{kasprzak,yamamoto,snoke}, clear evidence of the non-equilibrium nature of the polariton condensate has been observed in the ballistic outward polariton flow from small-sized condensates~\cite{maxime_small,noi_shape} as well as in the spontaneous appearance of vortices in the presence of a significant disorder potential~\cite{vortices}.
 
Several theoretical papers have recently investigated the effect of the non-equilibrium condition on the elementary excitation spectrum~\cite{littlewood,nonresonant}, and in particular have pointed out the diffusive nature of the Goldstone mode at low wavevectors.
In what follows, we shall adopt the point of view of our previous paper~\cite{nonresonant} where the dynamics of a non-equilibrium condensate was discussed at the mean-field level in terms of a generalized non-equilibrium Gross-Pitaevskii equation.

\begin{figure}[ht]	\includegraphics[width=0.8\columnwidth,angle=0,clip]{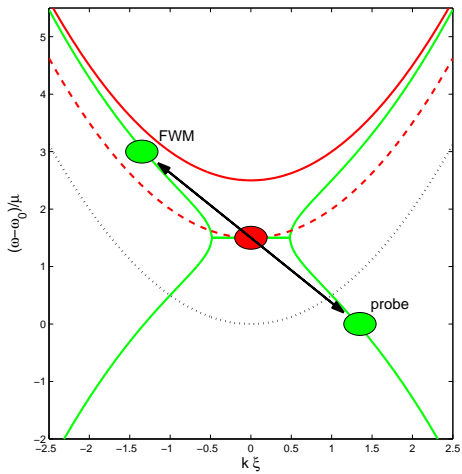} 
\caption{Plot of the elementary excitation spectrum  in a spatially homogeneous geometry and sketch of the proposed FWM experiment.
The red dot indicates the wavevector $k=0$ and the energy $\omega_c$ of the condensate.
The solid green line is the dispersion \eq{eq:spectr} of the elementary excitations on top of the condensate. The arrows show the parametric scattering process on which our proposed scheme is based: the condensate is probed at the lower green dot and the response is probed via the emission at the upper green dot. 
The dotted black line is the dispersion of free polaritons at linear regime. 
The solid and dashed red lines include the blue-shift due to condensate-condensate and condensate-reservoir interactions as mentioned in the text.
System parameters: $\gamma=\Gamma=g\,n_c$, $g_R n_R=0.5 g\,n_c$.}
	\label{fig:sketch}
\end{figure}

This gives the following form for the dispersion of the elementary excitations on top of a spatially homogeneous non-equilibrium condensate:
\begin{equation}
\omega_\pm(k)-\omega_c =-{i\Gamma}/{2}\pm \sqrt{[\omega _{Bog}(k)]^2 -
{\Gamma^{2}}/{4}}\; \label{eq:spectr}
\end{equation}
which is plotted in Fig.\ref{fig:sketch} as a green solid line.
In the low-$k$, diffusive region of the spectrum the $\pm$ signs correspond to respectively the phase and the density branches. In the high-$k$ region where the standard equilibrium Bogoliubov dispersion is recovered, they instead correspond to the normal and anomalous ``ghost'' branches. In the following of the paper, a specific attention will be paid to the ``ghost'' branch which is a signature of coherent polariton-polariton interactions.

The condensate energy $\omega_c$ (red dot) is blue-shifted with respect to the bottom of the lower polariton branch (black dotted line) by the condensate-condensate and condensate-reservoir interactions, $\omega_c = \omega_0 + g n_c + g_R n_R$.
Here, $n_{c}$ and $n_R$ are the densities respectively of the condensate and of the exciton reservoir, while the $g$ and $g_R$ coupling constants characterize the interactions respectively between a pair of condensate polaritons and between condensate polaritons and reservoir excitons.
The Bogoliubov spectrum of equilibrium condensates has the standard form
 $\omega_{Bog}(k)=[\varepsilon_k(\varepsilon_k+2\,g\,n_c)]^{1/2}$
in terms of the bare polariton dispersion at linear regime. 
Within the parabolic approximation, this can be written as $\varepsilon_k=\hbar k^{2}/2m_{LP}$.
As discussed in~\cite{nonresonant}, the effective width $\Gamma$ is defined as
$	\Gamma=\alpha\, \beta\, \gamma/(1+\alpha\beta)$,
in terms of the scaled pump intensity $\alpha=P/P_{th}-1$ above the threshold value $P_{th}$ and the $\beta$ coefficient defined in terms of gain rate $R(n_R)$ of the condensate from the excitonic reservoir as $\beta=n_{R}^{0}R^{\prime }\left( n_{R}^{0}\right) /R\left(n_{R}^{0}\right)$.
Well above the threshold $\alpha\gg 1$, the effective width $\Gamma$ recovers the empty-cavity polariton decay rate $\gamma$. 

Note how the dispersion of elementary excitations significantly differs from the naive Hartree prediction $\omega=\omega_0+ 2 g_R n_R + g n_c+ \hbar k^2/2m$ (dashed red line) and rather recovers in the high-momentum region the value $\omega=\omega_0+ g_R n_R + {\mathbf 2} g n_c+\hbar k^2/2m$ which correctly includes a factor $2$ due to bosonic exchange (solid red line).
Provided the effect of saturation~\cite{ciuti_review} on the polariton interactions is negligible, these simple arguments can be used to isolate the contribution of respectively the condensate-reservoir and the condensate-condensate interactions to the blue-shift, and possibly even evaluate the fraction of condensed excitons with respect to the total number of excitons in the microcavity.

\section{The four wave mixing spectroscopy scheme  \label{sec:FWM}}

Even though a simple luminescence experiment should in principle be able to reveal both the positive and the negative ``ghost'' branch of the elementary excitation spectrum shown in Fig.\ref{fig:sketch}, experimental observations have not been able to provide unambiguous evidence of the ``ghost'' branch yet~\cite{kasprzak,snoke,yamamoto_bog}: most likely the very weak luminescence coming from the ``ghost'' branch is hidden by the much stronger emission from the condensate and the upper branch.

The physical process underlying our proposal is sketched in Fig. \ref{fig:sketch}. 
Elementary excitations are created on top of the condensate by injecting extra polaritons with a probe laser beam at a finite in-plane wavevector $\kk$ tuned at a frequency $\omega$. 
The response of the system is then observed via the coherent light emission at an opposite wavevector $-\kk$ at an energy $2\omega_c-\omega$: the existence of a coherent coupling between the symmetrically located frequencies $\omega$ and $2\omega_c-\omega$ and wavevectors $\pm\kk$ stems from the fact that the elementary excitations of the condensate consist of a coherent superposition of plane waves at $(\kk,\omega)$ and $(-\kk,2\omega_c-\omega)$~\cite{BECbook,castin_houches}.
Equivalently, this same process can be interpreted in nonlinear optical terms as a stimulated parametric scattering where a pair of condensate polaritons are coherently scattered into one more probe polariton plus one FWM polariton of symmetric wavevector and frequency. Within the standard language of four-wave mixing, two of the three input beams are provided by the condensate itself, and only the third comes from the incident laser.

\subsection{Transmittivity}
\label{sec:T}

The same linearization procedure that was used in~\cite{nonresonant} to obtain the dispersion of the elementary excitations can be straighforwardly used to determine the response of the system to a (weak) probe laser of amplitude $E_{pr}(\kk,\omega)$. The simplest quantity to consider is the probe transmission.

\begin{figure}[ht]
\begin{center}
\includegraphics[width= \columnwidth,angle=0,clip]{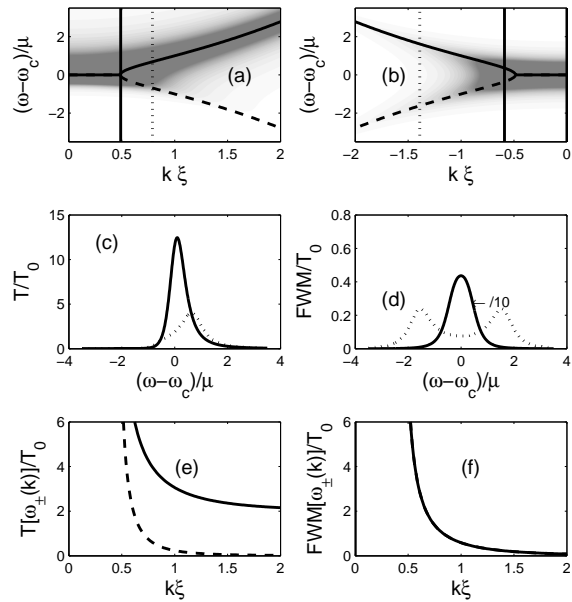} 
\end{center}
\caption{Transmittivity (a) and FWM (b) spectra of a non-equilibrium condensate excited with a probe laser at wave vector $k>0$ and frequency $\omega$. 
The full and dashed green lines indicate the positive and negative Bogoliubov branches from Eq. \eq{eq:spectr}. Cuts of the transmission and FWM spectra at given $k$ values are shown in panels (c) and (d); the corresponding $k$ values are indicated by the vertical lines in panels (a,b). 
The peak value of the transittivity and the FWM signal on the respectively the normal and ghost Bogoliubov branches are shown in (e) and (f) as solid and dashed lines. In (f) the two quantities coincide. All spectra are normalized to the empty-cavity, linear regime transmittivity $T_0$.
System parameters: $\Gamma=\gamma=g\,n_c$.
}
\label{fig:spectrfwmtr}
\end{figure}

In the usual approximation that the effective reservoir relaxation rate $\gamma_R$ is much larger than all other frequency scales in the problem, the reservoir dynamics can be eliminated and the amplitude of the transmitted field has the form:
\begin{equation}
	E_{tr}(\kk,\omega) = \frac{\sqrt{T_0}\,\gamma}{2}\,\frac{g n_c + \varepsilon_\kk +\omega +i\Gamma/2}{(\omega-\omega_+(\kk))(\omega-\omega_-(\kk))} E_{pr}(\kk,\omega).
	\label{eq:respu}
\end{equation}
The frequencies $\omega_\pm(\kk)$ that appear in the resonant denominators were defined in \eq{eq:spectr}: the transmission is resonantly enhanced when the the probe is on resonance with an elementary excitation of the polariton condensate, either on the positive energy branch or on the negative energy, ghost branch.
A complete density plot of the transmittivity $T=|E_{tr}/E_{pr}|^2$ in the $(\omega,k)$ plane is shown in Fig.\ref{fig:spectrfwmtr}(a). For the sake of simplicity, the transmission $T$ is normalized to the linear regime resonant transmittivity $T_0$ of the unloaded cavity.

At large momenta $k$, the intensity of the ghost mode is much weaker than the one of the upper, normal mode and quickly tends to zero. Already for $k\xi=1$, the ghost resonance is extremely weak and almost invisible in the cut shown as a dotted line in Fig.\ref{fig:spectrfwmtr}(c). 
On the other hand, the normal resonance tends to the unloaded cavity peak of height $T/T_0=1$ and width $\gamma$.
For decreasing values of $k$, the intensities of the two resonances get closer and eventually stick at the singular point $k\xi = 0.5$ of the dispersion [Fig.\ref{fig:spectrfwmtr}(e)]. 

In the low-$k$ diffusive region where the two branches are degenerate $\textrm{Re}[\omega_\pm]=0$, the transmission intensity significantly exceeds the peak transmittivity $T_0$ of the unloaded cavity [solid line in Fig.\ref{fig:spectrfwmtr}(c)]: this gain feature is due to the condensate, that is able to amplify the probe beam. 
A similar phenomenology was found and discussed in~\cite{goldstone} for the case of an optical parametric oscillator.

While the appearance of the ghost branch in the transmission spectra of Fig.\ref{fig:spectrfwmtr} is a signature of the presence of the condensate, the transmittivity on the normal branch can exceed the linear regime value $T_0$ even for pump intensities well below the condensation threshold. 
Even if not sufficient to overcome losses, some gain is in fact present also in this case~\cite{paul}, as witnessed by the polariton linewidth decreasing below the unloaded cavity value $\gamma$.

\subsection{FWM signal}

The same method can be used to obtain an expression for the amplitude of the four wave mixing signal:
\begin{multline}
E_{FWM}(\kk,\omega) =- \frac{\sqrt{T_0}\,\gamma}{2}\,\frac{g n_c+i\Gamma/2}{(\omega-\omega_+(\kk))(\omega-\omega_-(\kk))} 
\times \\ \times E^*_{pr}(-\kk,2\omega_c-\omega)\; .
\label{eq:respv}
\end{multline}
It is interesting to note that both the collisional ($\propto g n_c$) and gain saturation ($\propto \Gamma$) nonlinearities contribute to the FWM signal.
The response function $S_{FWM}=|E_{FWM}/E_{pr}|^2$  is plotted in the upper-left panel of Fig.\ref{fig:spectrfwmtr}. Thanks to the symmetry under the exchange $\omega\rightarrow -\omega$, the heights of the two peaks corresponding to the normal and the ghost branches are equal.

This is a crucial advantage of the FWM technique as compared to the transmission spectroscopy discussed in Sec.\ref{sec:T} or the resonant Rayleigh scattering discussed in~\cite{cambridge}. Differently from this latter scheme, FWM does not rely on the presence of a disorder potential and therefore is not affected by speckle-like modulations of the detected signal.
As it happened for the probe transmission, also the FWM signal results strongly enhanced in the diffusive region and may eventually become stronger than the probe itself. On the other hand, it decreases quite quickly for larger values of $k$.

Note that both direct polariton-polariton collisions and nonlinear gain saturation effects contribute to the FWM signal via respectively the terms $g\,n_c$ and $i\Gamma/2$ in the numerator of \eq{eq:respv}: some FWM signal is therefore expected to appear even for negligible polariton-polariton collisions, but is in this case peaked very close to the free polariton dispersion.
On the other hand, the presence of a FWM signal appears to be a conclusive evidence of the presence of a coherent condensate: the same calculation below threshold would in fact give a vanishing coherent FWM amplitude in the weak probe limit. Gain saturation is in fact not active in this regime and FWM processes involving the exciton reservoir only provide an incoherent background without any specific resonant feature.

\subsection{Reflectivity}

As many microcavity samples are grown on an absorbing substrate, transmission measurements are not always possible. While the FWM expression \eq{eq:respv} is the same in both reflection and transmission geometries, the interference between direct reflection of light on the external mirror and the secondary emission from the cavity makes the expression of the reflection amplitude a bit more complicate than the transmission amplitude \eq{eq:respu}:
\begin{equation}
	E_{r}(\kk,\omega) = \left(1 - i \frac{\gamma_{ph}}{2} \frac{g n_c + \varepsilon_\kk +\omega +i\Gamma/2}{(\omega-\omega_+(\kk))(\omega-\omega_-(\kk))} \right) E_{pr}(\kk,\omega).
	\label{eq:refl}
\end{equation}
Note that the linewidth $\gamma_{ph}$ that appears at the numerator of \eq{eq:refl} is the linewidth of the cavity-photon mode in the absence of any excitonic resonance. This quantity is to be distinguished from the linewidth $\gamma$ of the polariton branch that appeared in the previous formulas.
In terms of the Hopfield coefficients $U_{ph,x}^{LP}$ that quantify the cavity-photonic and the excitonic contents of the lower polariton branch, the polariton $\gamma$, exciton and cavity-photon $\gamma_{x,ph}$ linewidths are related by~\cite{ciuti_review}
\begin{equation}
\gamma=\gamma_x\,|U^{LP}_x|^2+\gamma_{ph}\,|U^{LP}_c|^2.
\end{equation}

\begin{figure}[ht]
	\begin{center}
	\includegraphics[width= \columnwidth,angle=0,clip]{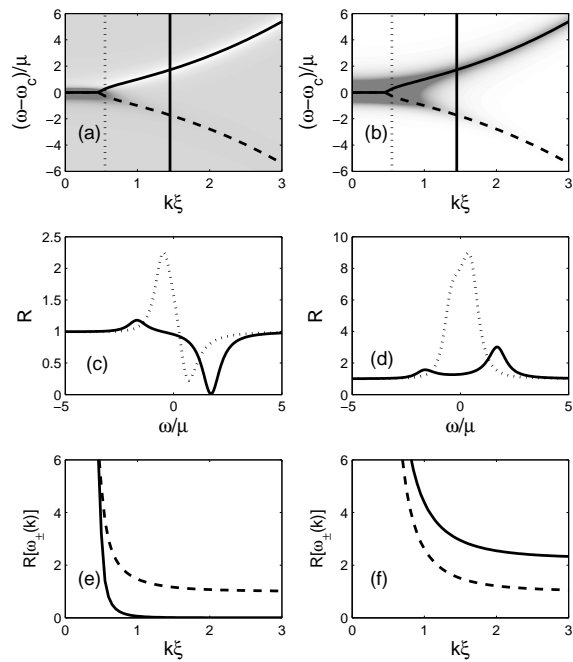} 
\end{center}
	\caption{Panels (a,b): $(k,\omega)$ plot of the reflectivity for $\gamma = \Gamma =g\,n_c$ (a) and $\Gamma = 0.4 \gamma=g n_c$ (b). (c,d) show reflection spectra at fixed values of $k$ as indicated by the vertical lines in panels (a) and (b). 
The peak values of the reflectivity on the normal and ghost Bogoliubov branches are respectively shown in (e) and (f) as solid and dashed lines. 
Parameters: $\gamma=\gamma_{ph}=\gamma_x$.}
	\label{fig:spectrefl}
\end{figure}

As it involves subtle interference effects, the reflection spectrum turns out to be sensitive to the ratio of the linewidth of excitations in the empty and pumped cavity $\gamma/\Gamma \geq 1$, as well as to the ratio between the cavity-photon and the polariton linewidths $\gamma_{ph}/\gamma$. 
Here we focus our attention to the experimentally most relevant case where $\gamma_{ph}=\gamma_x=\gamma$.
Examples of spectra are shown in Fig. \ref{fig:spectrefl} for different values of the pumping intensity. Panels (a,c) refer to a case well above threshold for which $\Gamma/\gamma\simeq1$, while panels (b,d) are made for a lower (but still above threshold) value of the pump intensity at which $\Gamma/\gamma=0.4$.
 
In panel (a), a minimum in the reflectivity is visible on resonance with the positive Bogoliubov branch, while the reflected intensity on resonance with the negative branch exceeds the incident intensity.
As for the transmission and the FWM signal, the amplification effect is the strongest in the diffusive region at low $k$ [panel (c)]. At high values of $k$, the dip corresponding to the normal branch remains fully visible, while the peak corresponding to the ghost branch disappears. This once again confirms that the FWM mixing scheme is the most suited tool for the detection of the ghost branch.

The same quantities are plotted in panels (b,d,f) for a smaller value of $\Gamma/\gamma=0.4$, i.e. for a pump intensity closer to the threshold. As a consequence of the narrower resonance line, the reflected beam is now more intense than the incident one on both the normal and the ghost branches. The amplification in the low-$k$ diffusive region is also stronger than in the case $P\gg P_{th}$ considered before.

\section{Probing the elementary excitation spectrum in finite size and/or disordered condensates \label{sec:probe}}

Even in the absence of trapping potentials, the spatial extension of polariton condensates is generally limited by the size of the laser spot that is used to pump the microcavity~\cite{noi_shape}. Typically, state of the art polariton condensates have a typical size of the order of tens of microns and may show a considerabe inhomogeneous broadening of their spectral features as a consequence of the inhomogeneous density profile. Although this effect can be reduced by choosing top-hat pump beams, the interpretation of FWM experiments is the clearest if one focusses the probe spot onto a small region at the center of the condensate where the density is almost flat and one spatially selects the emission from the same central region.

\begin{figure}[ht]
	\begin{center}
	\includegraphics[width= \columnwidth,angle=0,clip]{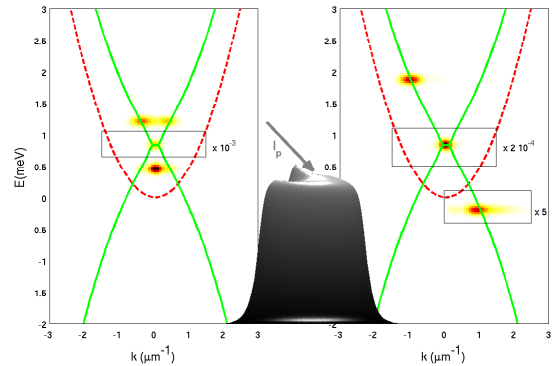} 
\end{center}
	\caption{(Color online) Numerical results for the $(k,\omega)$ emission pattern from a probed polariton condensate in a transmission geometry. The condensate frequency is $\omega_c=0.83$ meV and probe laser has a $1\mume$ diameter. The central wave vector and the frequency of the probe are $k_p = 0.6\,\mume^{-1}$, $\omega_p=0.3$ meV (left panel) and $k_p = 1.5\,\mume^{-1}$, $\omega_p=-0.2$ meV (right panel). 
The full and dashed lines show the analytical prediction the excitation spectrum and the linear regime polariton dispersion. The central surface plot shows the real space polariton density profile corresponding to the right hand panel. In all panels, the condensate is excited with a top hat laser with a $30\,\mume$ diameter and only the emission originating from a central region with a $20\,\mume$ diameter is recorded in the left and right panels. The finite frequency spread of the emission lines originates from the finite time window of about 12 ps from which the spectrum was extracted.}
	\label{fig:respnum}
\end{figure}

The result of numerical calculations based on the full polariton Gross-Pitaevskii equation~\cite{nonresonant} is summarized in Fig.\ref{fig:respnum}. The case of a probe beam of diameter $1\,\mume$ much smaller than the full condensate size and tuned at a frequency slightly below the condensate frequency $\omega_c$ is considered.

The case of a quite high probe wavevector $k_p=1.5\mume^{-1}$ tuned close to resonance with the ghost branch is shown in the right panel of the figure. As the $k$ distribution is narrow as compared to the central wavevector $k_p$, the response closely resembles the one of the spatially homogeneous case discussed in the previous section. 
A single FWM response peak appears around $-k_p$ and with a $k$-space linewidth narrower than the one of the incident probe: only those wavevector components that are on resonance with the ghost branch effectively contribute to the FWM signal.

A remarkable feature of the ghost branch is visible in the real space density profile of the probed condensate shown in the central panel of the figure:  as the group velocity of the ghost branch is opposite to the wavevector $\kk$, the perturbation due to the probe laser (with a central wavevector $k_p>0$ pointing to the right) propagates through the condensate in the leftwards direction and concentrates on the left-hand side of the probe spot.

The case of a probe laser with a smaller wave vector and a frequency closer to, but still below the condensate frequency is shown in the left panel. This choice dramatically modifies the qualitative shape of the response: in particular, note how the maximum of the response at the probe frequency lies between the two excitation branches of the homogeneous system, a spectral region where collective modes are strongly affected by the finite size of the system and possibly by the non-trivial spectral shape of the condensate~\cite{noi_shape,footnote}. Although information on the excitation modes of the condensate is hardly obtained in this configuration, the very presence of a FWM signal is still a direct proof of the presence of a coherent condensate.

\begin{figure}[ht]
	\begin{center}
	\includegraphics[width= \columnwidth,angle=0,clip]{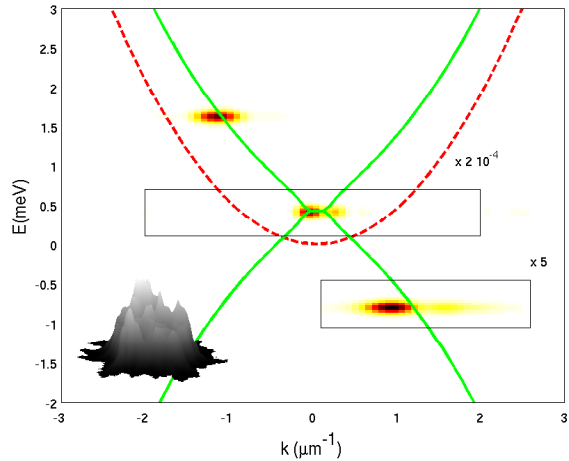} 
\end{center}
	\caption{(Color online) Emission pattern for the same configuration as in the right hand panel of Fig.~\ref{fig:respnum}, but in the presence of a disorder potential acting on the polaritons. The height of the disorder potential is in the 1 meV range. Inset: density profile of the disordered condensate.}
	\label{fig:dis}
\end{figure}

The situation is a bit more complex in the presence of disorder. As any experimental sample is inevitably far from being perfectly homogeneous, it is important to briefly summarize the main features that we observe in our numerical simulations when a disorder potential acting on polaritons is included. A complete study is postponed to further work.

The case of a weak disorder is illustrated in Fig.~\ref{fig:dis}: the strength of the disorder potential is chosen to be strong enough to substantially modulate the polariton density (see inset), but weak enough not to split the condensate into several frequency components~\cite{syncro}. 
In this case, the FWM signal remains clearly visible: the wavevector distribution is strongly broadened by the disorder, but the main features that were observed in the right panel of Fig.~\ref{fig:respnum} are still apparent.

The situation is dramatically different whenever multiple condensates are present with a substantial spatial overlap. In this case, the main effect of the probe beam is to redistribute the intensity among the different condensate frequencies and no clear FWM signal is easily identified among the many spectral lines forming the emission spectrum.

\section{Conclusions \label{sec:concl}}

We have proposed and analyzed a four-wave mixing spectroscopy scheme to probe the excitation spectrum of a non-equilibrium polariton condensate. A clear evidence of the negative frequency `ghost' branch is predicted to appear in the FWM spectra. The presence of this feature depends crucially on the coherent nature of the condensate and can therefore be used as an additional probe of the condensate coherence. Mapping out the FWM resonance as a function of frequency and wave vector gives a precise access to the elementary excitation spectrum. We have numerically demonstrated that a spatially localized probe can be used to overcome inhomogenous broadening effects by restricting the measurement to the central region of the condensate where the density is almost flat.

\section{Acknowledgements}

We wish to thank V. Savona, D. Sarchi, M. Richard, K. Lagoudakis,  A. Baas and B. Pietka for continuous stimulating discussions on polariton condensates. IC acknowledges financial support from the italian MIUR, the EuroQUAM-FerMix program, and the french IFRAF.


\begin{thebibliography}{99}
\bibitem{footnote} This interpretation is supported by the numerical observation that the response is again peaked on the Bogoliubov excitation branch if a larger condensate is considered.
\end{thebibliography}
\end{document}